\RequirePackage{amsmath}
\documentclass[runningheads,a4paper]{llncs}
\usepackage{amssymb}
\usepackage{graphicx}
\usepackage{url}
\usepackage{mathrsfs}
\usepackage{bm}
\usepackage{hyperref}
\usepackage[caption=false]{subfig}
\usepackage{algpseudocode}
\usepackage{algorithm}
\usepackage{acronym}
\usepackage{dsfont}
\usepackage{multirow}

\newcommand{\keywords}[1]{\par\addvspace\baselineskip

\noindent\keywordname\pagenumbering{gobble}\enspace\ignorespaces#1}

\begin{document}

\mainmatter  

\title{Hawkes processes for credit indices time series analysis: How random are trades arrival times?}
\titlerunning{Hawkes processes for credit indices time series analysis}

%
%
\author{Achraf Bahamou%
\and Maud Doumergue \and Philippe Donnat 
}
\authorrunning{Bahamou, A., Doumergue, M., Donnat, P.}

\institute{Hellebore Capital Ltd\\
Michelin House, London\\}

%
%

\toctitle{Modelling credit indices time series: How random are trades arrival times?}
\tocauthor{Authors' Instructions}

\maketitle

\begin{abstract}
    Targeting a better understanding of credit market dynamics, 
    the authors have studied a stochastic model named Hawkes process. 
    Describing
    trades arrival times, this kind of model allows for the capture of self-excitement and mutual interactions phenomena.
    The authors propose here a simple yet conclusive method for fitting multidimensional Hawkes processes with exponential kernels, based on a maximum likelihood non-convex optimization.
    The method was successfully tested on simulated data, then used on new publicly available real trading data for three European credit indices, thus enabling quantification of self-excitement as well as volume impacts or cross indices influences.


\end{abstract}


\section{Introduction}

\subsection{From credit derivative indices to Hawkes processes}

\begin{figure}
    \centering
    \includegraphics[height=4cm]{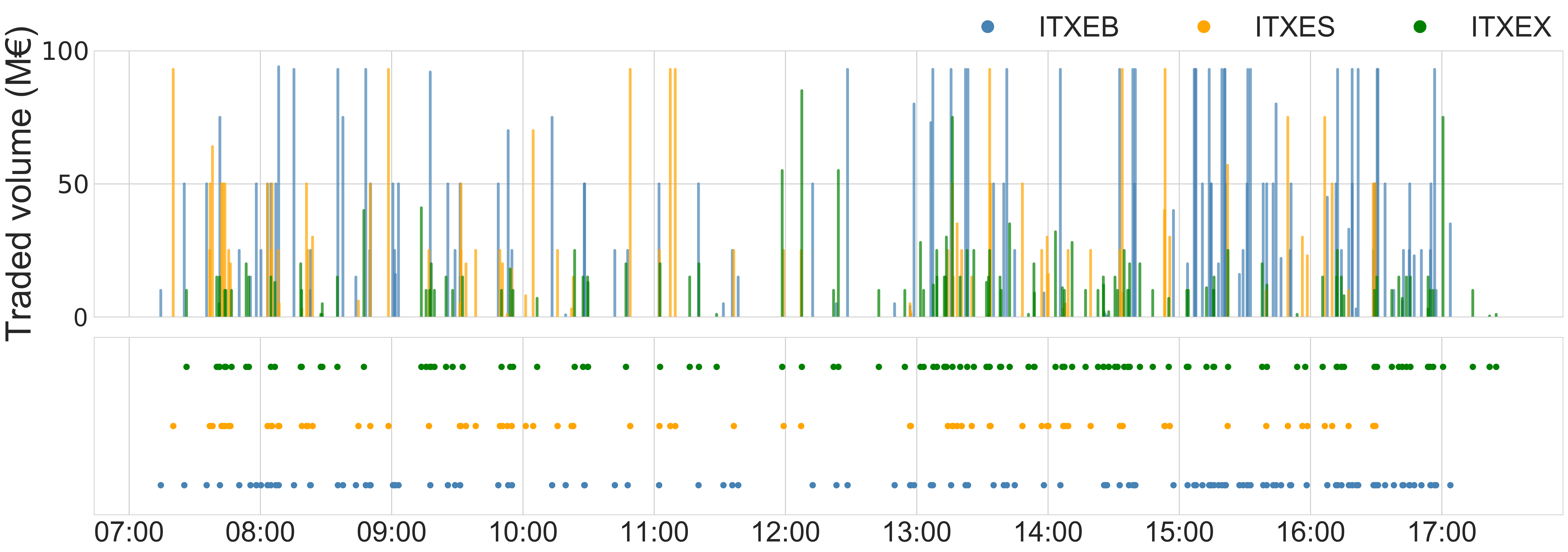}
    \caption{Publicly reported trading activity on 22/05/2018 (\textit{times}, \textit{volumes}) for European credit indices \textsc{itxeb}, \textsc{itxes} \& \textsc{itxex}. 
    }
    \label{fig:1}
\end{figure}

Credit indices are financial instruments comprised of a set of credit securities, mainly used to hedge credit default risk. Each new index series is issued every 6 months and expires after a defined maturity. Though liquid, those indices are traded at a rather "mid-frequency" rate, with trades occurring at a minute scale.
For European indices, the market is continuously open from 7:00 to 17:00, with a total of five to ten billion euros reported each day on average. Fig~\ref{fig:1} gives an idea of the activity on a sampled day of trading. If recent regulations have led to a greater public reporting of trading activities, thus releasing more amount of traded data, this data is by essence quite sparse. Picturing such market behaviour over time represents a challenging opportunity: it originally motivated the work presented in this article.

This study focuses on three principal European credit indices, of the most traded 5-year maturity\footnote{Data provided by \href{otcstreaming.com}{otcstreaming.com}, and available on demand.
}: the \textit{Main Index}, noted here \textsc{itxeb}, a combination of 125 equally weighted investment grade entities; the \textit{Crossover Index}, noted \textsc{itxex}, with 75 sub-investment grade names; and the \textit{Senior Financial}, noted \textsc{itxes}, a subset of 30 financial entities from Main Index, referencing senior debt.

\vspace{-20pt}

\begin{figure}
    \centering 
    \includegraphics[height=3.1cm]{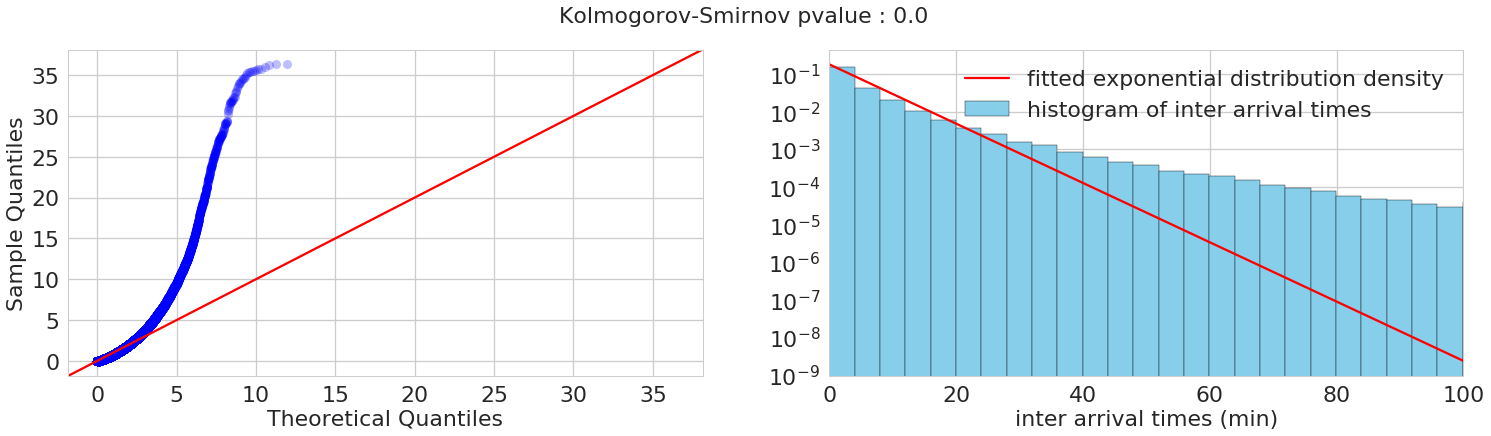}%
    \caption{\textsc{itxeb}, from 15/01/2017 to 15/05/2018: \textit{right}: log-scaled distribution of inter-arrival times; \textit{left}: Q-Q plot. The \textit{red lines} represent the best exponential law fitting. With a p-value of 0.0, this model is clearly inappropriate.
    }
    \label{fig:2}
\end{figure}

When trying to capture the underlying dynamics of this market, one of the first questions that comes to mind is: how random is the timing of trades?
A very naive approach is to model trades arrival times as purely random and uncorrelated processes, for example a Poisson process by trying to fit an exponential law over the distribution of inter-arrival times. Yet, such a model fails to fit the data: from Fig~\ref{fig:2}, we can notice a high density of very short inter-arrival times, suggesting that some self-excitation phenomenon occurs, with trades triggering more trades; an intuition that every practitioner of the field would confirm.

\vspace{-10pt}
\begin{table}
\centering
\begin{tabular}{lcccccc}
            & $\mathbb{E}(\Delta t)$ & $\sigma(\Delta t)$ & $Q_1(\Delta t)$ & $Q_2(\Delta t)$ & $Q_3(\Delta t)$& Number of trades \\ 
{\textsc{itxeb}} & 7~min                       & 10~min                 & 1~min                & 3~min                & 8 min        & 22~227                                                                       \\ 
{\textsc{itxex}}& 8~min                      & 12~min                  & 2~min                & 4~min                & 10 min & 18~152                                                                               \\ 
{\textsc{itxes}}& 17~min                      & 24~min                   & 3~min               & 9~min                & 21 min    & 7~194                                                                             \\ 
\end{tabular}
\caption{Statistics on inter-arrival times $\Delta t$
(period: 03/01/2017 to 14/12/2017)}
\label{table1}
\end{table}
\vspace{-10pt}

\vspace{-20pt}
\vspace{-10pt}

Literature review reveals that one model has recently driven considerable attention and a growing interest from both the scientific and quantitative communities: Hawkes processes. 
This model, introduced in 1971 by Hawkes (\cite{hawkes0}, \cite{hawkes1}) to characterize earthquake tremors, can be used to describe timing of trades and their cross influences from a point process perspective.

After summarizing related work for Hawkes processes, 
we will recall the theoretical framework around this stochastic model.
We will then focus on  optimization methods for fitting such a process and describe Two Stage Hawkes Likelihood Optimization (2SHLO), a maximum likelihood based algorithm to fit multidimensional Hawkes processes with a parametric exponential kernel. 
Lastly, we will present our results, firstly on simulated data, then 
on credit derivative indices mentioned above. Hawkes processes in finance have mainly been applied to high frequency data. Is such a model appropriate for "mid-frequency" credit data? Can we specify the impact of volumes in trades or describe mutual influences between different indices? 






\subsection{Related work}

Hawkes processes were originally designed for seismology analysis and  deeply studied for that purpose by Ogota \cite{ogata1}, \cite{ogata78}, \cite{ogata81}. The concept has been utilized to model other effects where self and cross ignitions happen, such as social media tweets cascading \cite{socialmedia}, crime occurrences due to gang retaliations \cite{crime}, or financial market events. As regards to financial applications, Bowsher \cite{bowsher} proposed in 2002 a generalized Hawkes process model taking into account night gaps, inter-day dependencies as well as intraday seasonality (with a piece-wise baseline intensity), which was used to model interactions between trades and price changes and also estimate of the price volatility based on mid-quote intensity for NYSE stock. Market price microstructure \& market impact (influence of market orders on forthcoming prices) was modelled in \cite{bacry2}, price impact was also studied in \cite{papanicolaou}. \cite{bacry3} focused on the impact of volume and order types on the limit order book.  \cite{bacry1} (2015) exposes a full review of Hawkes process applications to finance, such as price \& volatility modeling, market reflexivity measurement, order book modeling or risk contagion modeling.

\section{Modeling Self and Cross Excitement with Hawkes Processes}

This section re-frames the required formalism, as very clearly stated in \cite{bacry1}, \cite{laub} or \cite{toke}, starting from point and counting process definitions, through the key concept of intensity function. Hawkes model formulation is detailed for the reader, with a peculiar focus on exponential kernel structure. 




\subsection{Core concepts: from counting \& point processes to intensity functions}

\begin{definition}[Point Process]
    Let $(\Omega, \mathscr{F}, P)$ be a probability space. \\
    Let $(t_k)_{k \in \mathbb{N}^{*}}$ be a sequence of non-negative random variables such that $\forall \ k \in \mathbb{N}^{*}$, $t_k < t_{k+1}$. $(t_k)_{k \in \mathbb{N}^{*}}$ is called a (simple) point process on $\mathbb{R}_+$.
\end{definition}


\begin{definition}[Counting Process]
    Let $(t_k)_{k \in \mathbb{N}}$ be a point process. The right-continuous stochastic process defined for all $t \in \mathbb{R}_+$ as $N(t) = \sum_{k \in \mathbb{N}^{*}} 1_{t_k \leq t}$ is called the counting process associated with $(t_k)_{k \in \mathbb{N}^{*}}$.
\end{definition}


The study of point processes goes through a single mathematical object, the \textit{intensity function}, which is defined as the conditional probability density of occurrence of an event in the immediate future.

\begin{definition}[Intensity Function]
    Let $N$ be a counting process adapted to a filtration $\mathscr{F}_t$. The left-continuous intensity is heuristically defined as 
    \begin{equation}
        \lambda (t | \mathscr{F}_t ) = \lim_{h \rightarrow 0} \mathbb{E} \bigg[ \frac{N(t+h) - N(t)}{h} \bigg| \mathscr{F}_t \bigg]
    \end{equation}
\end{definition}

The intensity function depends on the choice of filtration $\mathscr{F}_t$, which represents the amount of information available until time $t$ (See \cite{daley} for a rigorous definition of the intensity function). In the context of Hawkes processes, we will simply use the natural filtration, with all previous information being available, and consider $\lambda (t)$.

\paragraph{Homogeneous Poisson process}

\

One of the simplest point processes is the homogeneous Poisson process, for which the intensity function is constant over time: $\forall \  t \geq 0, \ \lambda(t) = \lambda$.
In that case, durations (or inter-event waiting times) are independent and identically distributed (following an exponential distribution of hazard rate $\lambda$). As presented in the introduction, credit trades cannot be modeled by this memory-less model. Let's introduce Hawkes processes, a peculiar kind of non-homogeneous Poisson processes with linear dependencies over functions of past events.



    



\subsection{Self-excitement: one-dimensional Hawkes processes}




\begin{definition}[One-dimensional Hawkes Processes]

    Let $(t_k)_{k \in \mathbb{N}}$ be a point process and $N$ the associated counting process, such that its intensity function $\lambda$ is defined for each time $t \geq 0$ as 
    
    \begin{equation}
    \lambda(t) = \mu(t) + \int_{-\infty}^{t} \phi(t-\tau) \ {dN}(\tau)  = \mu(t) + \sum_{t_i < t}  \phi(t-t_i)
    \end{equation}
    
    
    where $\mu : \mathbb{R} \mapsto \mathbb{R}_+$ is an exogenous base intensity and $\phi : \mathbb{R}_{+} \mapsto \mathbb{R}^{+}$
     is a non-negative, measurable function such that
     $||\phi||_1 = \int_0^{\infty}\phi(s)ds < 1$. 
     
    $N$ is called a Hawkes process with baseline $\mu$ and kernel $\phi$.

\end{definition}

The kernel $\phi$ expresses the positive influence of past events on the current value of the intensity.
Each jump  $dN(\tau) \ne 0$ increases the probability of
future events through the kernel $\phi$.
Clustering effects and branching structure are well depicted by Hawkes processes, with the baseline activity generating immigrant  events and descendant events enhancing the intensity. 

For this study, we focus on \textit{exponential kernels}, defined as  $\phi(t) = \alpha \beta \ e^{- \beta t} \ \mathds{1}_{t \geq 0 }$, with parameters $\alpha, \beta \geq 0 $, which allows easy interpretation: $\alpha$ or \textit{adjacency} represents the weight of previous events while $\beta$ is the \textit{decay} / typical duration of influence for a past event.

The \textit{branching ratio}  $ n = ||\phi||_1 = \int_0^{\infty}\phi(s)ds $ represents the average number of descendants for any event. For exponential kernel,  $n=\alpha$. 
\subsection{Including mutual-excitement with multidimensional Hawkes processes}

\begin{definition}[Multidimensional Hawkes Processes]

    Let $M \in \mathbb{N}^{*}$ and $\{(t_k^i)_k\}_{i=1,...,M}$ be a M-dimensional point process. 
    
    We denote by $N_t = (N_t^1,..., N_t^M)$ the associated counting process such that its vector intensity function  $\lambda : \mathbb{R}_{+} \mapsto \mathbb{R}_+^M$,  is defined as, for all $t \geq 0$, $i \in {1, ..., M} $: 
        
    \begin{equation}
    \lambda_{i}(t)=\mu_i+\sum_{j=1}^{M}\int_0^t \phi_{i,j}(t-\tau)d N_j(\tau)=\mu_i+\sum_{j=1}^{M}\sum_{n=1}^{N_j(t)} \phi_{i,j}(t-t_{j,n}),
    \end{equation}
    
    with $\mu = (\mu_i)_{i=1,...,M}$ 
    an exogenous base intensity vector
    
    and 
    $\phi(t) =  (\phi_{i,j})_{i,j=1,...,M}$ a matrix-valued kernel that is component-wise positive and causal (null values when $t<0$) and with each component belonging to the space of $L^1$-integrable functions.

    $N_t$ is called a multidimensional (or multivariate)  Hawkes process with baseline $\mu$ and kernel $\phi$.
\end{definition} \label{multiD}

The choice of an exponential kernel can be generalized for multivariate Hawkes processes with kernel components being expressed as $\phi_{i,j}(t) = \alpha_{i,j} \beta_{i,j} e^{-\beta_{i,j} t} \mathds{1}_{t  \geq 0}$.

This article focuses on Hawkes processes with exponential kernel and constant baselines, to which we propose a maximum likelihood based fitting method.

\section{Hawkes Processes: Model Calibration}

The most commonly used technique for parametric inference of Hawkes processes is a direct numerical optimization of the Maximum Likelihood, which was first introduced in the work of Ogata (1978) \cite{ogata78}. He proved the asymptotic consistency and efficiency of the Maximum Likelihood Estimator (MLE) under the assumption of stationary condition of the underlying point process. The negative log-likelihood function of a multidimensional Hawkes process over the time interval $[0,t]$ is given in Daley \& Vere-Jones \cite{daley} Proposition 13.1.VI by :

\begin{equation}  \label{eq:56}
\mathcal{L}_t(\bm{\lambda}):=-\sum_{i=1}^{M}\left(\int_{0}^t\log\lambda_i(\tau)d N_{i}(\tau)-\int_0^t\lambda_i(\tau)d\tau\right).
\end{equation}

Depending on the shape of kernels, $\mathcal{L}_t$ is generally non-convex. Classic non-convex numerical optimization pitfalls are to be feared: a direct numerical optimization could converge to a merely local minimum.
On the other hand, it may take too many iterations to converge as the negative log-likelihood function can be flat on some regions of the space of parameters. This problem is also faced when using an Expectation Maximization (EM) algorithm to find the MLE.




On the computational side, the repeated evaluation of the negative log-likelihood can be highly 
time consuming, essentially due to the nested sum in the first part of \ref{eq:56} where the conditional intensity is also expressed as a sum over the history. 

\subsection{Maximum Likelihood Estimation (MLE) for exponential multidimensional Hawkes processes}



Let's consider a M-dimensional multivariate Hawkes model with constant baselines $\mu_i$ and kernel functions $\phi_{i,j}$ as defined in \ref{multiD}. The choice of an exponential kernel 
has proved to be very interesting as 
it allows intuitive and meaningful interpretations of its parameters and also reduces the computational cost of the evaluation of the negative log-likelihood function as noted by Ogata (1981) \cite{ogata81} who exhibited a recursive formula that eliminated the nested sum evaluation problem.

The existing methods to fit exponential parameters through MLE 
directly use non-linear optimization algorithms such as Nelder-Mead (also called downhill simplex method) or BFGS, with performance decreasing as the number of parameters increases. This represents quite an issue as $M(2M+1)$ parameters describe a M-dimensional Hawkes process with exponential kernel.

In addition, other existing methods often make the strong assumption that the decays $\beta_{i,j}$ of the exponential kernel are given and fixed a priori. Consequently, the estimation of each kernel function is equivalent to the estimation of its adjacency coefficient $\alpha_{i,j}$ and since the conditional intensity $\lambda_i(t)$ is linear with respect to kernel functions $\phi_{i,1}(t),\ldots,\phi_{i,p}(t)$, the negative log-likelihood function $\mathcal{L}_t$ is convex with respect to all parameters. Therefore we can use the widely available convex optimization machinery to find the global minimum efficiently. The main drawback of this method is the possible inaccuracy of the decays parameters $\beta_{i,j}$ which can lead to model mismatch.

\subsection{Introducing Two Stage Hawkes Likelihood Optimization (2SHLO):}
In order to combine the benefits of both approaches described above, we propose the following estimation method to fit exponential Hawkes processes.

The negative log-likelihood function for exponential multivariate Hawkes processes is given by (see Ogata\cite{ogata81} for the recursive formulation):

\begin{multline}
    \mathcal{L}_t(\bm{\mu},\bm{\alpha},\bm{\beta}) = \sum_{m=1}^{M}\bigg(\mu_mT \ + \ \sum_{n=1}^{M} \alpha_{mn} \sum_{\{ k:t_k^n < T\}}[ 1-e^{-\beta_{mn}(T-t_k^n)}]  \\
    - \sum_{\{ k:t_k^n < T\}} log[\mu_m + \sum_{n=1}^{M}\alpha_{mn}\beta_{mn}\sum_{\{ k:t_k^n < t_i^m\}} e^{-\beta_{mn}(t_i^m-t_k^n)}] \bigg)
\end{multline}




As we mentioned earlier, if $\beta$  is fixed, then $\mathcal{L}_t(\cdot,\cdot,\beta)$ is convex with respect to parameters $(\bm{\mu},\bm{\alpha})$. As a result it can be minimized using any convex optimization algorithm, for example an Accelerated Gradient Descent (AGD) 
which converges to a global minimum $(\bm{\mu}^*(\bm{\beta}),\bm{\alpha}^*(\bm{\beta}))$ such that:
\begin{align}
     \mathcal{L}_t^*(\bm{\beta}) = \mathcal{L}_t(\bm{\mu}^*(\bm{\beta}),\bm{\alpha}^*(\bm{\beta}),\bm{\beta}) = min_{(\bm{\mu},\bm{\alpha})} \mathcal{L}_t(\bm{\mu},\bm{\alpha},\bm{\beta})
\end{align} \label{eqmin}

$\mathcal{L}_t^*(\bm{\beta})$ is still a non-convex function but defined on a space with a lower dimension than the initial parameter space. Any non-linear heuristic algorithms can now be used to minimize $\mathcal{L}_t^*(\bm{\beta})$ leading to the MLE as\footnote{if we call $m = min_{(\bm{\mu},\bm{\alpha},\bm{\beta})} \mathcal{L}_t(\bm{\mu},\bm{\alpha},\bm{\beta})$, the following inequality holds for all  $(\bm{\mu},\bm{\alpha},\bm{\beta})$:
$
    m \leq \mathcal{L}_t(\bm{\mu}^*(\bm{\beta}),\bm{\alpha}^*(\bm{\beta}),\bm{\beta}) \leq \mathcal{L}_t(\bm{\mu},\bm{\alpha},\bm{\beta})
$
\ref{eqmin} follows immediately by taking the minimum over $(\bm{\mu},\bm{\alpha},\bm{\beta})$ in each side of the inequality. 
}:
\begin{align}
    min_{(\bm{\mu},\bm{\alpha},\bm{\beta})} \mathcal{L}_t(\bm{\mu},\bm{\alpha},\bm{\beta}) =
    min_{(\bm{\beta})}min_{(\bm{\mu},\bm{\alpha})} \mathcal{L}_t(\bm{\mu},\bm{\alpha},\bm{\beta}) 
\end{align}



\noindent We summarize the fitting method as follow : 
\vspace{-10pt}
\begin{algorithm}[H]
  \caption{Two Stage Hawkes Likelihood Optimization (2SHLO)}

  \begin{algorithmic}
    \State Start from mean of inter arrival times $\bm{\beta_0}$  
    \For  {each step of Nelder-Mead method until convergence} 
        \For  {each needed computation of $\mathcal{L}_t^*(\bm{\beta_i})$} 
        \State Start from a random $(\bm{\mu_0},\bm{\alpha_0})$ 
        \State Use Accelerated Gradient Descent to optimize $\mathcal{L}_t(\bm{\mu},\bm{\alpha},\bm{\beta_i})$
        
        \State Retrieve resulting $\mathcal{L}_t^*(\bm{\beta_i})$
        \EndFor 
    \EndFor  
    \State \Return  last $(\bm{\beta}, \  \bm{\mu}^*(\bm{\beta}),\ \bm{\alpha}^*(\bm{\beta})) $
  \end{algorithmic}
\end{algorithm}
\vspace{-10pt}

\subsection{Goodness of fit}

The \textit{compensator function} $\Lambda$ of a point process with intensity $\lambda$ is defined as 
$\forall \  t \geq 0 , \ \Lambda(t) = \int_{0}^{t} \lambda (s| \mathscr{F}_t) ds $. To assess the goodness of fit of our estimation, we use the residual point process analysis theorem, as stated in \cite{daley}: 

\textit{The transformed sequence $\{t^*_k\} = \{\Lambda (t_k)\} $ is a realization of a unit rate Poisson process if and only if the original sequence  $\{t_k\}$ is a realization from the point process defined by $\lambda$.}



	

 The goodness of fit can be tested by comparing the transformed estimated times to a standard Poisson process using Q–Q plots and comparison of density functions.

\section{Results - 2SHLO in practice}




We have been running our experiments in Python 3, relying on \href{https://x-datainitiative.github.io/tick/#}{\textit{tick}}\cite{tick} library for Hawkes process simulations, AGD optimization and analytic tools. 2SHLO is also benefiting of well-known \textit{scipy} scientific package.

To provide an idea of the computational performance of the fitting method, we mention that, on average, it takes 4.69 seconds to fit 2 dimensional exponential hawkes process with training length of T = 10000.

\subsection{Validating 2SHLO performances over simulated data}

To generate Hawkes processes, a few methods are available (as summarized in \cite{bacry1}), either based on thinning, time-change or cluster algorithm. We have here used \textit{tick} functions for simulations which are based on the thinning algorithm described in (Ogata, 1981, p.25, Algorithm 2) \cite{ogata81}.

To validate the performance of the fitting algorithm, we ran a series of 100 simulation and fitting procedure on the simulated 2D Hawkes time-stamps using different end times T to have different training sizes and using the following simulation parameters :

\[
\mu = \bigl(\begin{smallmatrix}
0.1 \\ 
0.2 
\end{smallmatrix}\bigr) \quad 
\alpha = \bigl(\begin{smallmatrix}
0.5 & 0.00 \\ 
0.4 & 0.3 
\end{smallmatrix}\bigr) \quad 
\beta = \bigl(\begin{smallmatrix}
0.3 & 0.00 \\ 
0.2 & 0.2 
\end{smallmatrix}\bigr)
\]

\begin{figure}
    \centering
    \subfloat{{\includegraphics[height=5.5cm]{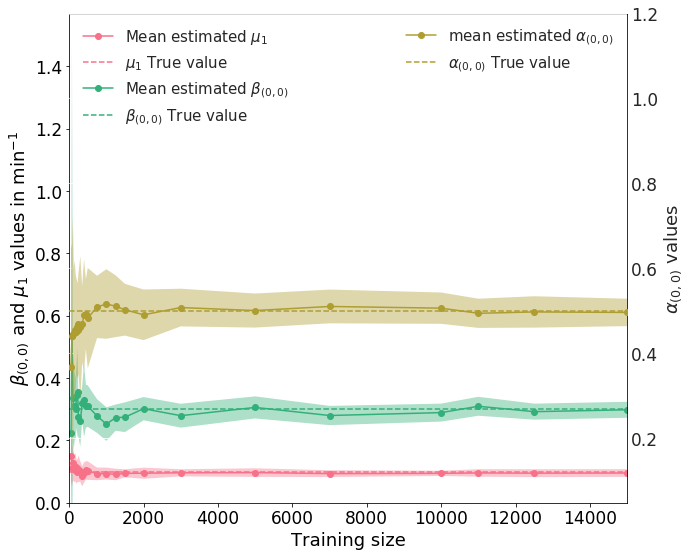}}}
    \caption{Estimated parameters of a 2D exponential Hawkes process trained with 2SHLO on simulated datasets with increasing lengths}%
    \label{fig:fittingparams}%
\end{figure}

Fig.~\ref{fig:fittingparams} confirms that the estimated parameters converge to their true optimal values for increasing N number of ticks in the training set. This reassures that when the Hawkes process is well specified and indeed recovers the optimal parameters asymptotically. 

\subsection{Calibrating univariate Hawkes processes on "mid-frequency" credit derivative trades}

We have fitted a modified one-dimensional Hawkes process on the period from 15/01/2017 to 15/12/2017 of reported trades data\footnote{The trading data have been cleaned by discarding transactions that do not reflect market signals such as roll and switch trades} on 3 indices \textsc{ITXEB}, \textsc{ITXES} and \textsc{ITXEX}. The modification consists of imposing a null value of the intensity in the gap between two days and the standard exponential kernel format during trading days so as to account for the absence of trading activity overnight. The model is thus slightly modified and described in the appendix section in a more detailed way. Before making this modeling choice, we experimented with a more elaborate model that accounts for the overnight spillover effect as described in the work of C. G.Bowsher \cite{bowsher} where the author defines the intensity as recursively depending on the level of the stochastic parts of the intensity at the end of the previous trading day and thus extending the effect of trading days over time. The fitting procedure resulted into a null parameter estimation of the spillover effect which motivated the choice of the model modification described above. 

\begin{figure}
    \centering
    \includegraphics[height=8cm]{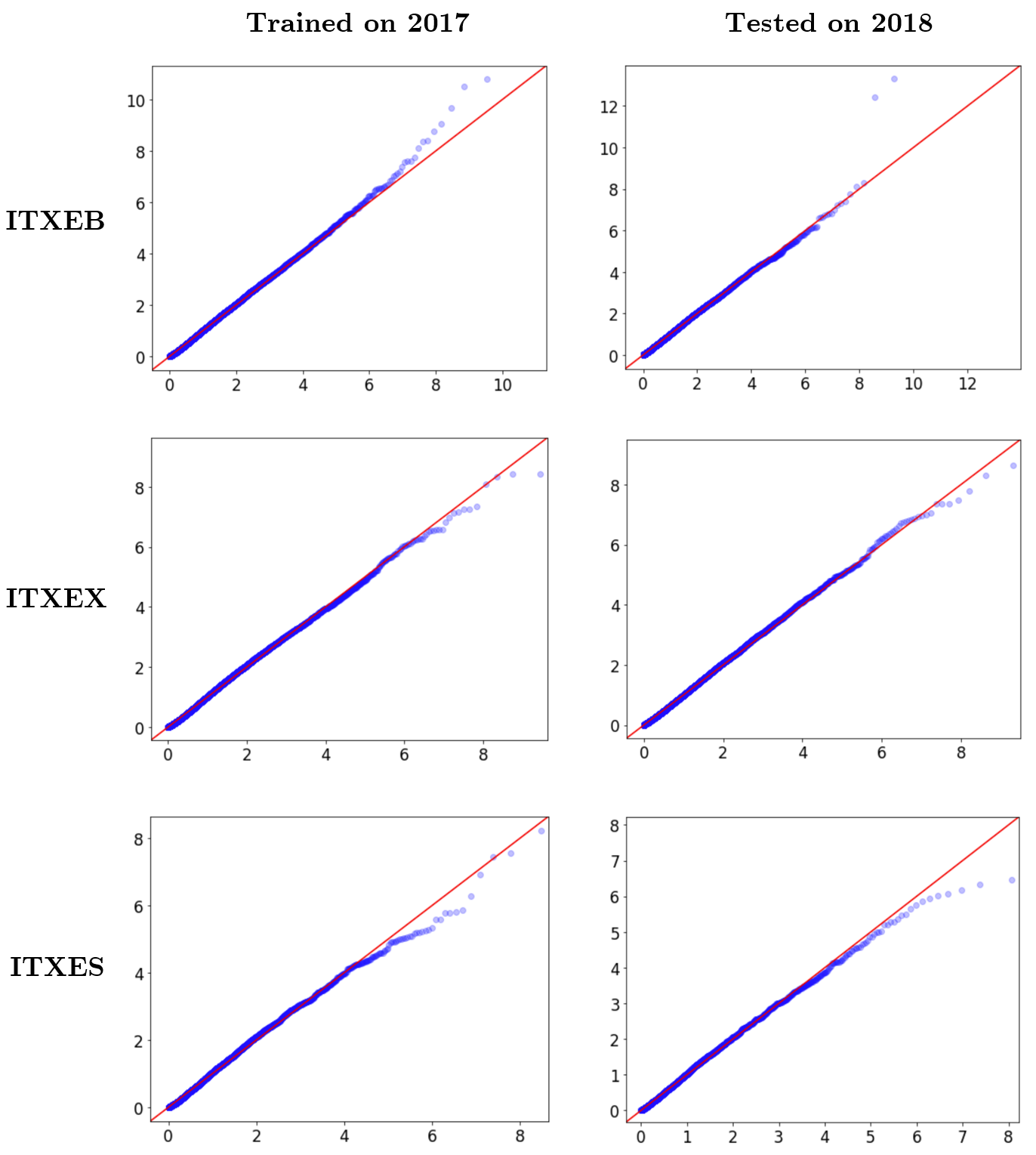}%
    \caption{Hawkes process fitted on 2017 data and tested on 2018 data for each index}
    \label{fig:indicestests}
\end{figure}

To validate the 'universality' of our estimates, we tested the goodness of fit of the Hawkes model with the estimated parameters on out of sample period from 15/01/2018-15/07/2018. The model fitted well for the 3 indices time series (see Fig.~\ref{fig:indicestests}) suggesting that each series has stable intrinsic parameters that characterize the pattern of trades arrival times.

\begin{table}
\centering
\begin{tabular}{l|l|l|l|l}
\cline{2-4}
                            & Estimated $\mu^{-1}$ & Estimated $\alpha$ & Estimated $\beta^{-1}$ &  \\ \cline{1-4}
\multicolumn{1}{|l|}{ITXEB} & 22 min                                          & 0.62                   & 20 min                                      &  \\ \cline{1-4}
\multicolumn{1}{|l|}{ITXEX} & 24 min                                          & 0.60                   & 23 min                                       &  \\ \cline{1-4}
\multicolumn{1}{|l|}{ITXES} & 58 min                                          & 0.65                   & 36 min                                       &  \\ \cline{1-4}
\end{tabular}
\caption{Estimated parameters for \textsc{ITXEB}, \textsc{ITXES} and \textsc{ITXEX} over the period (2017-01-15 to 2017-12-15)}
\label{table:1dhawkes}
\end{table}

\subsection{Measuring the influence of traded volumes}

\begin{figure}
    \centering
    \subfloat{{\includegraphics[height=2.9cm]{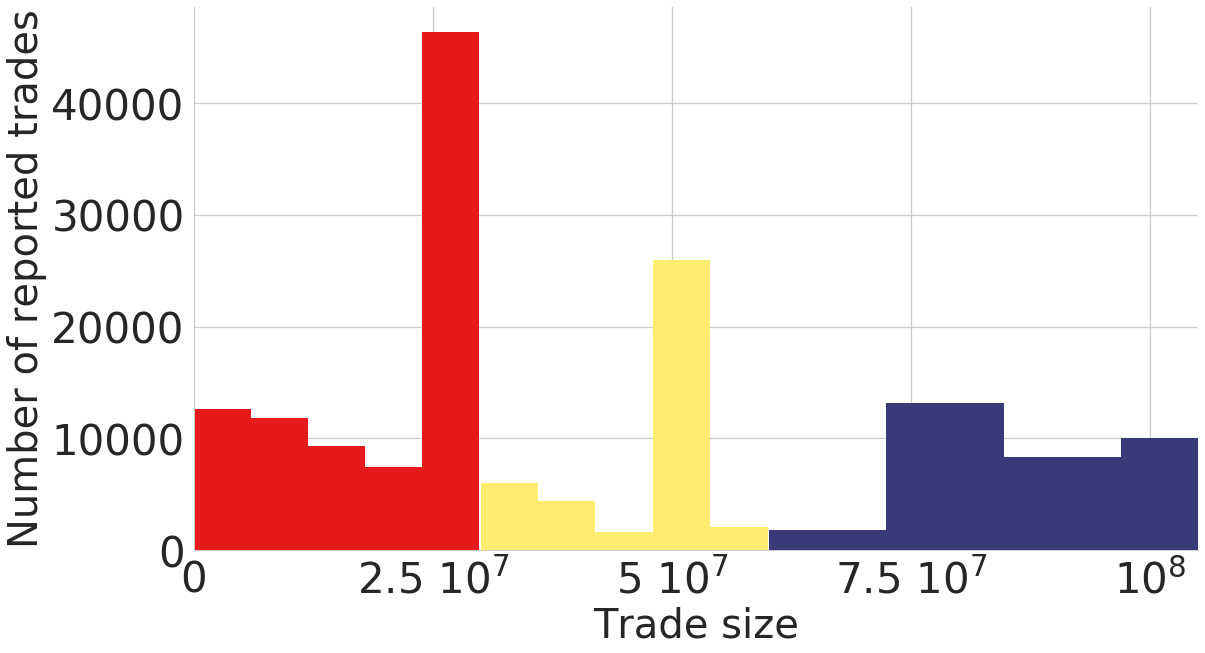} }}%
    \qquad
    \subfloat{{\includegraphics[height=3.5cm]{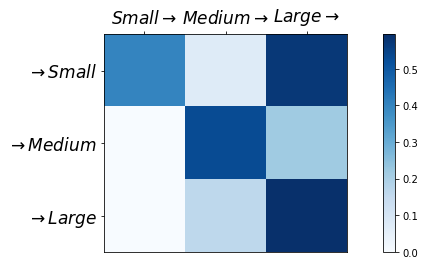} }}%
    \caption{Distribution of trade sizes over one week trading period 2018-05-21:2018-05-25 (\textit{left}; adjacency matrix $(\alpha_{i,j})$ of 3D exponential Hawkes fitted on 3 volume bin series (for index \textsc{itxeb}). }%
    \label{fig:sizeinflu}%
\end{figure}

When considering the distribution of trade volumes for credit indices, volume clusters are clearly outlined. For example, for index \textsc{itxeb}, three bins can be distinguished: "small" trades, "medium" trades and "big" trades (See Fig.~\ref{fig:sizeinflu}). Using multivariate Hawkes processes model, volume impacts have been modeled by considering 3 counting processes, splitting trades per volume size.

By using the same model as the last section to take overnight gaps into consideration, the fitting algorithm trained on 2018 data resulted in the following parameters estimates ($\mu$ and $\beta$ expressed in minute$^{-1}$) : 

\[
\frac{1}{\mu} = \bigl(\begin{smallmatrix}
40 \\ 
42 \\ 
38
\end{smallmatrix}\bigr) \quad 
\alpha = \bigl(\begin{smallmatrix}
0.40 & 0.07 & 0.57\\ 
0.00 & 0.53 & 0.21\\ 
0.00 & 0.16 & 0.59
\end{smallmatrix}\bigr) \quad 
\frac{1}{\beta} = \bigl(\begin{smallmatrix}
12 & 21 & 53\\ 
71 & 18 & 12\\ 
53 & 14 & 34
\end{smallmatrix}\bigr)
\]

Visualizing in Fig.~\ref{fig:sizeinflu} the branching ratio which corresponds the the values of $\alpha$ allows some intuitive interpretation of the fitted model as it shows that large trades are purely self exciting process and they have a major influence in triggering small trades, we also notice that all trade categories have a non negligible self excitation component.

\subsection{Measuring cross interactions between traded indices}

We have adopted again a multivariate Hawkes processes with null intensity overnight model to study the cross interaction between indices trades times. The fitting algorithm trained on 2017 data resulted in the following parameters estimates ($\mu$ and $\beta$ expressed in minute$^{-1}$) :

\[
\frac{1}{\mu} = \bigl(\begin{smallmatrix}
29 \\ 
30 \\ 
125
\end{smallmatrix}\bigr) \quad 
\alpha = \bigl(\begin{smallmatrix}
0.44 & 0.25 & 0.20\\ 
0.23 & 0.37 & 0.18\\ 
0.07 & 0.08 & 0.45
\end{smallmatrix}\bigr) \quad 
\frac{1}{\beta} = \bigl(\begin{smallmatrix}
23 & 9 & 21\\ 
14 & 13 & 31\\ 
39 & 17 & 25
\end{smallmatrix}\bigr)
\]

\begin{figure}
    \centering
    \includegraphics[height=3cm]{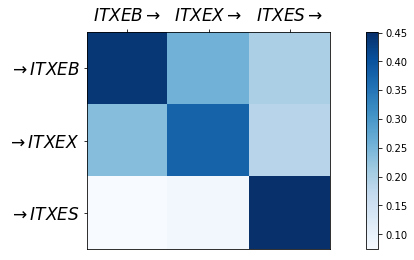}
    \caption{adjacency matrix $(\alpha_{i,j})$ of 3D exponential Hawkes fitted on 3 indices series (\textsc{itxeb}, \textsc{itxes}, \textsc{itxex}) for the period 2017-01-15 - 2017-12-15.}
    \label{fig:8}
\end{figure}

From Fig.~\ref{fig:8} 
we can notice that \textsc{itxes} index is the least influenced by the other indices with the very low cross excitation components and a high self excitation component which was expected due the its low liquidity. We also notice that the estimated baselines of all indices are greater for this multivariate model compared to the uni-dimensional model which can be explained by the fact that we decreased the 'Poissonnian' behaviour of the arrival times by including more information about the influence of other indices. 

\section{Conclusion}

Through point process analysis, Hawkes processes allows a relatively simple and interpretable representation of time dependant events with self and mutual excitements.
With 2SHLO; a maximum likelihood based algorithm mixing convex and non-convex optimization, we have been able to fit fastly and properly exponential kernel multivariate Hawkes processes.
The algorithm has been applied to credit derivative trading data, an unexplored "mid-frequency" market for such processes. We have been able to emphasize self excitement for three index trades and to measure impacts of traded volumes. With the same framework, we were also able to quantify cross influences between indices.
Next prospects turns towards testing 2SHLO performances on high dimensional datasets, for instance studying both price \& volume trade influences as well as market bid/ask dynamics on credit derivatives in a forecasting perspective.

\newpage

\section*{Appendix: Hawkes Processes Formulas}

\subsection{The Univariate Bowsher Hawkes process model}

\subsubsection{Definition:}

As described in \cite{bowsher}, the intensity of the univariate Bowsher process is defined recursively depending on the level of the stochastic parts of the intensity function at the end of the $(d-1)$th trading day and the contributions of the events occurring on day $d$. Thus it is necessary to perform the following data transformation: the real half-line is partitioned into intervals of length $l$ corresponding to the different trading days. This partition is written as :

\begin{align*}
    (0,\infty) = (0, \tau_1] \cup (\tau_1, \tau_2] \cup \dots \cup (\tau_{d−1}, \tau_d] ∪  \dots
\end{align*}

The model is thus defined by the (scalar) stochastic intensity :

\begin{align*}
    \forall t \geq 0, \quad \lambda(t) = \mu + \widetilde{\lambda}(t)
\end{align*}

such that :

\begin{align*}
    \forall t \in ]\tau_{d-1}, \tau_{d}] \quad
        \widetilde{\lambda}(t) = \pi\widetilde{\lambda}(\tau_{d-1})e^{-\rho(t-\tau_{d-1})} + \int_{[\tau_{d-1},t)}\alpha e^{-\beta(t-u)}dN(u)
\end{align*}

where the parameters used are :

\begin{itemize}
    \item $]\tau_{d-1}, \tau_{d}]$ is the interval defining the day number d
    \item $\mu$ is the baseline intensity
    \item $\pi$ is the spillover adjacency coefficient
    \item $\rho$ is the spillover decay
    \item $\alpha$ is the self excitement adjacency coefficient
    \item $\beta$ is the self excitement decay
    \item $dN$ is the process differentiate
\end{itemize}

\subsubsection{The Loglikelihood:}

The minus loglikelihood of a univariate point process is defined as :

\begin{align*}
    \mathcal{L}_T(\lambda) = -\left(
            \int_0^T \log \lambda(t) dN_i(t)
            - \int_0^T \lambda(t) dt
        \right)
\end{align*}

Computing the minus loglikelihood in the Hawkes Bowsher model results into the following closed form :

\begin{align*}
    \mathcal{L}_T(\lambda) = \mu T + \sum_{d \in days} \pi\widetilde{\lambda}(\tau_{d-1})(1-e^{-\rho(\tau_{d}-\tau_{d-1})}) \\
        + \alpha\sum_{t \in ]\tau_{d-1}, \tau_{d}]} (1-e^{-\rho(t-\tau_{d-1})}) - \sum_{t_i} \log \lambda(t_i)
\end{align*}

\subsubsection{The Fitting procedure:}

As the task of computing the gradient of each parameter of the loglikelihood is a fastidious one because of the recursive definition of the intensity, and also because we have only 5 parameters to fit $(\mu,\pi,\rho,\alpha,\beta)$, we choose to use the L-BFGS-B algorithm to perform the minimization of $\mathcal{L}_T(\lambda)$ which had delivered satisfying performance on fitting simulated data.

\subsubsection{Data simulation:}

We use the thinning algorithm to simulate artificial data described by the following algorithm :

\begin{algorithm}[H]
	\caption{Simulation by thinning}
	\begin{enumerate}
	\item Given Bowsher Hawkes process described as above 
		\item
		Set current time $T = 0$ and event counter $i = 1$
		
		\item
		While $i \leq N$
		\begin{enumerate}
			\item Set the upper bound of Poisson intensity $\lambda^*=\lambda(T)$ . \label{step:step-3a}
			\item Sample inter-arrival time: draw $u \sim U(0, 1)$ and let $\tau = -\frac{ln(u)}{\lambda^*}$ (as described in).
			\item Update current time: $T = T + \tau$. \label{step:step-3c}
			
			\item Draw $s \sim U(0, 1)$.
			
			\item
			If $s \leq \frac{\lambda(T)}{\lambda^*}$, accept the current sample: let $T_i = T$ and $i = i + 1$. \\
			Otherwise reject the sample, return to step (a). \label{step:step-3e}
			
		\end{enumerate}
	\end{enumerate}
	\label{alg:simulation-thinning}
\end{algorithm}

\subsection{The Day Gaps Multivariate Exponential Hawkes process model}

\subsubsection{Definition:}

The intensity of the Day Gaps Multivariate Exponential Hawkes process is defined just like the standard multivariate exponential Hawkes process model described in \cite{hawkes0} given by :

\begin{align*}
    \forall i \in [1 \dots D], \quad
        \lambda_i(t) = \mu_i + \sum_{j=1}^D \int \phi_{ij}(t - s) dN_j(s)
\end{align*}

Where :
\begin{itemize}
    \item $D$ is the number of nodes
    \item $\mu_i$ are the baseline intensities
    \item $\phi_{ij}$ are the kernels
    \item $dN_j$ are the processes differentiates
\end{itemize}

with an exponential parametrisation of the kernels:

\begin{align*}
    \phi_{ij}(t) = \alpha_{ij} \beta_{ij} \exp (- \beta_{ij} t) 1_{t > 0}
\end{align*}

\begin{itemize}
    \item $\alpha_{ij}$ are the adjacency of the kernel
    \item $\beta_{ij}$ are the decays of the kernel
\end{itemize}

With the modification consisting on considering the intensity as a null function between two consecutive days to take the day gaps into consideration :

\begin{align*}
    \forall t \text{ between two consecutive days,} \quad \lambda_i(t) = 0
\end{align*}

\subsubsection{The Loglikelihood:}

The minus loglikelihood of a Multivariate Point Process is defined as :

\begin{align*}
    \mathcal{L}_t(\bm{\lambda}):=-\sum_{i=1}^{M}\left(\int_{0}^t\log\lambda_i(\tau)d N_{i}(\tau)-\int_0^t\lambda_i(\tau)d\tau\right).
\end{align*}

Computing the minus loglikelihood in The Day Gaps Multivariate Exponential Hawkes process model results into the following closed form :

\begin{align*}
    \mathcal{L}_t(\bm{\lambda}):= D*\delta \sum_{m = 1}^M \mu_m + \sum_{m = 1}^M \sum_{n = 1}^M \alpha_{mn} \sum_{d \in days} \sum_{t_i^n \in \text{day d}} (1-e^{-\beta_{mn}(\tau_{d}-t_i^n)}) \\
        - \sum_{m = 1}^M \sum_{t_i^m} log(\mu_m+\sum_{n = 1}^M\alpha_{mn} \beta_{mn} R_{mn}(i))
\end{align*}

with $R_{mn}(i)$ are defined using the Ogoata \cite{ogata81} recursive formula :

\begin{align*}
    R_{mn}(i) &= \sum_{\{ k:t_k^n < t_i^m\}} e^{-\beta_{mn}(t_i^m-t_k^n)} \\
    &= e^{-\beta_{mn}(t_i^m-t_{i-1}^m)}R_{mn}(i-1)
        + \sum_{\{ k:t_{i-1}^m \leq t_k^n < t_i^m\}} e^{-\beta_{mn}(t_i^m-t_k^n)}
\end{align*}

Where :
\begin{itemize}
    \item $D$ is the number of days in the data
    \item $\delta$ is the length of one day
    \item $\mu_i$ are the baseline intensities
    \item $\alpha_{ij}$ are the adjacency coefficients
    \item $\beta_{ij}$ are the decays coefficients
    \item $\tau_{d}$ is the time defining the last time of the day number d
\end{itemize}

\subsubsection{The Fitting procedure:}

To fit this model, we use the algorithm described in this paper, using the projected Newton Descent for the convex optimization part and Nelder Mead for the non convex optimization step. The choice of the newton descent is justified by the easy computation of the gradient and hessian matrix which is sparse and because this algorithm has a quadratic convergence rate compared to gradient descent methods.

The gradient of $\mathcal{L}_t(\bm{\lambda})$ is thus given by the following close formulas :

\begin{align*}
    \frac{\partial \mathcal{L}_t(\bm{\lambda})}{\partial \mu _m} &= D*\delta-\sum_{t_i^m} \frac{1}{\mu_m+\sum_{n = 1}^M\alpha_{mn} \beta_{mn} R_{mn}(i)} \\
    \frac{\partial \mathcal{L}_t(\bm{\lambda})}{\partial \alpha _{mn}} &= \sum_{d \in days} \sum_{t_i^n \in \text{day d}} (1-e^{-\beta_{mn}(\tau_{d}-t_i^n)}) - \sum_{t_i^m} \frac{\beta_{mn} R_{mn}(i)}{\mu_m+\sum_{l = 1}^M\alpha_{ml} \beta_{ml} R_{ml}(i)}
\end{align*}

The hessian of $\mathcal{L}_t(\bm{\lambda})$ is given by the following close formulas :

\begin{align*}
     \frac{\partial^2 \mathcal{L}_t(\bm{\lambda})}{\partial \mu_m \partial \mu _m} &= \sum_{t_i^m} \frac{1}{(\mu_m+\sum_{n = 1}^M\alpha_{mn} \beta_{mn} R_{mn}(i))^2} \\
     \frac{\partial^2 \mathcal{L}_t(\bm{\lambda})}{\partial \mu_m \partial \alpha_{ml}} &= \sum_{t_i^m} \frac{\beta_{ml} R_{ml}(i)}{(\mu_m+\sum_{n = 1}^M\alpha_{mn} \beta_{mn} R_{mn}(i))^2} \\
     \frac{\partial^2 \mathcal{L}_t(\bm{\lambda})}{\partial \alpha_{ml} \partial \mu_m} &= \sum_{t_i^m} \frac{\beta_{ml} R_{ml}(i)}{(\mu_m+\sum_{n = 1}^M\alpha_{mn} \beta_{mn} R_{mn}(i))^2} \\
     \frac{\partial^2 \mathcal{L}_t(\bm{\lambda})}{\partial \alpha_{mk} \partial \alpha _{ml}} &= \sum_{t_i^m} \frac{\beta_{mk} R_{mk}(i)\beta_{ml} R_{ml}(i)}{(\mu_m+\sum_{n = 1}^M\alpha_{mn} \beta_{mn} R_{mn}(i))^2}
\end{align*}

and all other coefficients are null. The fitting algorithm is described as follow:

\begin{algorithm}[H]
  \caption{Two Stage Hawkes Likelihood Optimization using Projected Newton Descent}
  \begin{algorithmic}
    \State Start from mean of inter arrival times $\bm{\beta_0}$  
    \For  {each step of Nelder-Mead method until convergence} 
        \For  {each needed computation of $\mathcal{L}_t^*(\bm{\beta_i})$} 
        \State Start from a well chosen $(\bm{\mu_0},\bm{\alpha_0})$ 
        \State Use Projected Newton Descent to optimize $\mathcal{L}_t(\bm{\mu},\bm{\alpha},\bm{\beta_i})$ :
        
        \State Retrieve resulting $\mathcal{L}_t^*(\bm{\beta_i})$
        \EndFor 
    \EndFor  
    \State \Return  last $(\bm{\beta}, \  \bm{\mu}^*(\bm{\beta}),\ \bm{\alpha}^*(\bm{\beta})) $
  \end{algorithmic}
\end{algorithm}

And the projected newton descent is described as follow :

\begin{algorithm}[H]
  \caption{Projected Newton Descent}
  \begin{algorithmic}
    \State Start from a well chosen $x_0$   
    \For  {each step until convergence or iter max} 
        \State Set prev\_x = x, prev\_obj = obj
        \State Set step = 1
        \State Set shrink = 0.5 (user choice of a shrinx in $[0,1]$)
        \While{True} 
            \State next\_x = project(x - step*$(\nabla^2 f(x))^{-1} \nabla f(x)$)
            \State next\_obj = f(next\_x)
            \If{next\_obj $>$ prev\_obj}
                \State step *= shrink
            \Else 
                \State x = next\_x
            \EndIf
        \EndWhile
    \State test convergence abs(next\_obj - prev\_obj) / abs(prev\_obj) $<$ tolerance
    \EndFor
    
    \State \Return  x
  \end{algorithmic}
\end{algorithm}

\subsection{Some empirical results:}

\subsubsection{Loglikelihood landscape:}
Visualizing the minus loglikelihood function for 2D exponential Hawkes process with regards to two non convex variables $(\beta_{0,0},\beta_{1,1})$ resulted into the following figure, emphasizing the flatness of the function to minimize and the existence of valleys that pose problem to gradient based optimization :

\begin{figure}%
    \centering
    \includegraphics[height=5cm]{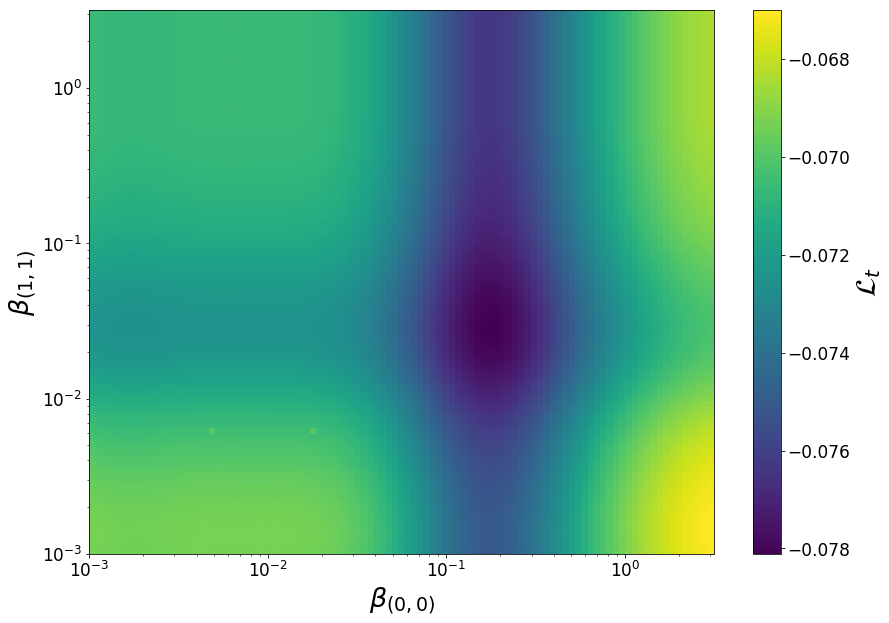} 
    \caption{$\mathcal{L}_t$ plot for a 2D exponential Hawkes process loglikelihood with respect to decays variables $(\beta_{0,0},\beta_{1,1})$, other parameters being fixed: $(\beta_{0,1},\beta_{1,0})=(0.1,0.1)$, $\mu=(0.02,0.25)$, $\alpha = ((0.3 , 0.15), (0.01 , 0.35))$}
    \label{fig:landscape4}
\end{figure}

\subsubsection{Goodness of fit size bins study:}
Assessing the goodness of fit of the multivariate Hawkes process fitted on 2017 \textsc{itxeb} reported trades split into three series with regards to the size of trades:

\begin{figure}[H]
    \centering
    \includegraphics[height=15cm]{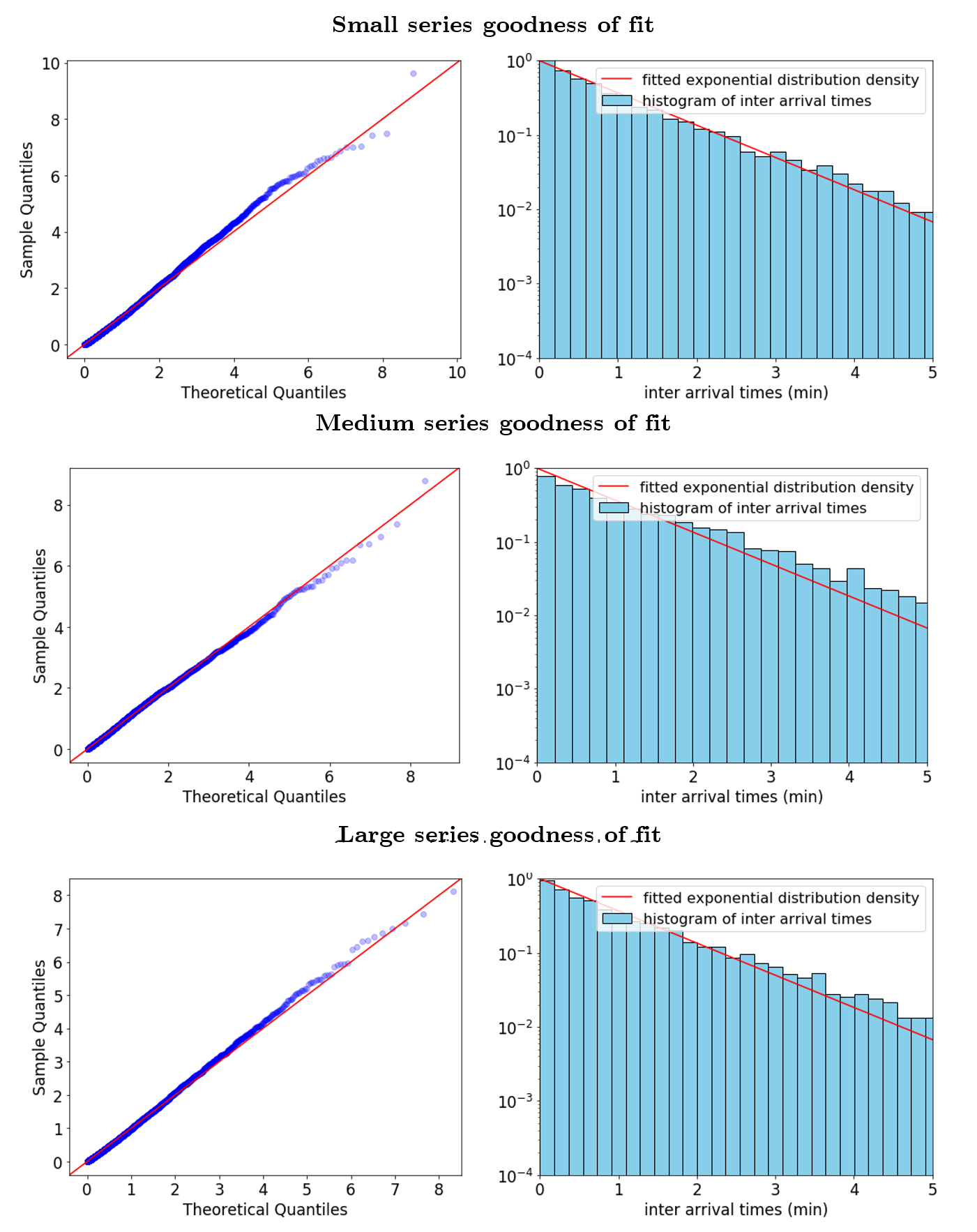} %
    \caption{QQPlots of goodness of fit of the multivariate Hawkes process fitted on 2017 \textsc{itxeb} by size bin reported trades }
    \label{fig:landscape5}%
\end{figure}

\subsubsection{Goodness of fit Multi index study:}
Assessing the goodness of fit of the multivariate Hawkes process fitted on 2017 \textsc{itxeb}, \textsc{itxes} and \textsc{itxex} reported trades 

\begin{figure}[H]
    \centering
    \includegraphics[height=15cm]{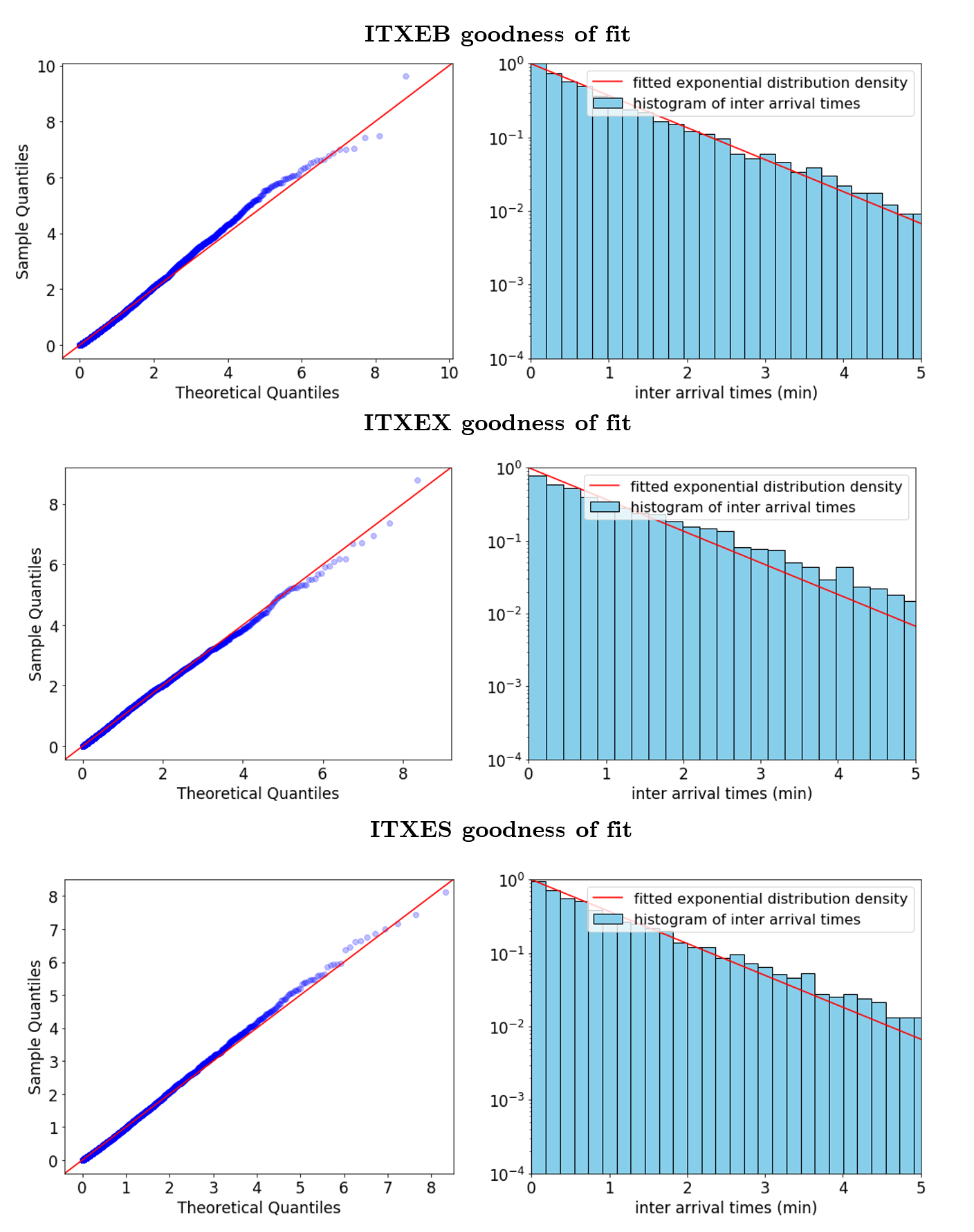} 
    \caption{QQPlots of goodness of fit of the multivariate Hawkes process fitted on 2017 \textsc{itxeb}, \textsc{itxes} and \textsc{itxex}}
    \label{fig:landscape6}%
\end{figure}

\end{document}